# DISTANCE'S QUANTIFICATION ALGORITHM IN AODV PROTOCOL

Meryem Saadoune[1] , Abdelmajid Hajami[2] and Hakim Allali[3]

Department of Mathematics and Computer Sciences, Hassan 1st University, Settat, Morocco

[1] saadoune.meryem@gmail.com
[2] abdelmajidhajami@gmail.com
[3] hakim-allali@hotmail.fr

## ABSTRACT

*Mobility is one of the basic features that define an ad hoc network, an asset that leaves the field free for the nodes to move. The most important aspect of this kind of network turns into a great disadvantage when it comes to commercial applications, take as an example: the automotive networks that allow communication between a groups of vehicles. The ad hoc on-demand distance vector (AODV) routing protocol, designed for mobile ad hoc networks, has two main functions. First, it enables route establishment between a source and a destination node by initiating a route discovery process. Second, it maintains the active routes, which means finding alternative routes in a case of a link failure and deleting routes when they are no longer desired. In a highly mobile network those are demanding tasks to be performed efficiently and accurately. In this paper, we focused in the first point to enhance the local decision of each node in the network by the quantification of the mobility of their neighbours. Quantification is made around RSSI algorithm a well known distance estimation method.*

## 1. INTRODUCTION

Mobile ad hoc network (MANET) is an appealing technology that has attracted lots of research efforts. Ad hoc networks are temporary networks with a dynamic topology which doesn't have any established infrastructure or centralized administration or standard support devices regularly available as conventional networks [1]. Mobile Ad Hoc Networks (MANETs) are a set of wireless mobile nodes that cooperatively form a network without infrastructure, those nodes can be computers or devices such as laptops, PDAs, mobile phones, pocket PC with wireless connectivity. The idea of forming a network without any existing infrastructure originates already from DARPA (Defense Advanced Research Projects Agency) packet radio network's days [2][3]. In general, an Ad hoc network is a network in which every node is potentially a router and every node is potentially mobile. The presence of wireless communication and mobility make an Ad hoc network unlike a traditional wired network and requires that the routing protocols used in an Ad hoc network be based on new and different principles. Routing protocols for traditional wired networks are designed to support tremendous numbers of nodes, but they assume that the relative position of the nodes will generally remain unchanged. In ad hoc, since the nodes are mobile, the network topology may change rapidly and unpredictably and the connectivity among the terminals may vary with time. However, since there is no fixed infrastructure in this network, each mobile node operates not only as a node but also as a router forwarding packets from one node to other mobile nodes in the network that are outside the range of the sender. Routing, as an act of transporting information from a source to a destination through intermediate nodes, is a fundamental issue for networks. [4]

The problem that arises in the context of ad hoc networks is an adaptation of the method of transport used with the large number of existing units in an environment characterized by modest computing capabilities and backup and fast topology changes.

According to the way of the creation and maintenance of roads in the routing of data, routing protocols can be separated into three categories, proactive, reactive and hybrid protocols. The pro-active protocols establish routes in advance based on the periodic exchange of the routing tables, while the reactive protocols seek routes to the request. A third approach, which combines the strengths of proactive and reactive schemes, is also presented. This is called a hybrid protocol.

Ad-hoc On-Demand Distance Vector routing protocol (AODV) [5] is a reactive routing protocol, who was standardized by the working group MANET [6] with IETF (Internet Engineering Task force), by the (RFC 3561).

The protocol's algorithm creates routes between nodes only when the routes are requested by the source nodes, giving the network the flexibility to allow nodes to enter and leave the network at will. Routes remain active only as long as data packets are traveling along the paths from the source to the destination .When the source stops sending packets, the path will time out and close.

In this paper we propose a solution that enables each node in the network to determine the location of its neighbors in order to create a more stable and less mobile road. For that purpose, we locally quantify the neighbor's distances of a node as the metric of mobility using AODV protocol.

The remainder of this paper is organized as follows. Section 2, describes briefly the AODV protocol. In Section 3, a summary of related work is presented. we present in Section 4 how to quantify, evaluate, estimate mobility in ad hoc network. Section 5 shows the algorithm used the quantification of the distance in AODV protocol. Section 6 presents some simulations and results. Finally Section 7 concludes this paper.

## 2. AD HOC ON-DEMAND DISTANCE VECTOR

AODV is an on-demand protocol which is capable of providing unicast, multicast [7], broadcast communication and Quality of Service aspects (QoS) [8], [9]. It combines mechanisms of discovery and maintenance roads of DSR (RFC 4728) [10] involving the sequence number (for maintains the consistency of routing information) and the periodic updates of DSDV [11].

At the discovery of routes, AODV maintains on each node transit information on the route discovery, the AODV routing tables contain:

- The destination address
- The next node
- The distance in number of nodes to traverse
- The sequence number of destination
- The expiry date of the entry of the table time.

When a node receives a packet route discovery (RREQ), it also notes in its routing table information from the source node and the node that just sent him the package, so it will be able to retransmit the response packet (RREP). This means that the links are necessarily symmetrical. The destination sequence number field of a route discovery request is null if the source has never been linked to the destination, else it uses the last known sequence number. It also indicates in this query its own sequence number. When an application sends a route discovery, the source waits for a moment before rebroadcast its search query (RREQ) road, after a number of trials, it defines that the source is unreachable.

Maintained roads is done by periodically sends short message application called "HELLO" , if three consecutive messages are not received from a neighbor, the link in question is deemed to have failed . When a link between two nodes of a routing path becomes faulty, the nodes broadcast packets to indicate that the link is no longer valid. Once the source is prevented, it can restart a process of route discovery.

AODV maintains its routing tables according to their use, a neighbor is considered active as long as the node delivers packets for a given destination, beyond a certain time without transmission destination, the neighbor is considered inactive. An entered routing table is considered active if at least one of the active neighbors using the path between source and destination through active routing table entries is called the active path. If a link failure is detected, all entries of the routing tables participating in the active path are removed.

## 3. RELATED WORK

In [12], a geometric mobility metric has been proposed to quantify the relative motion of nodes. The mobility measure between any pair of nodes is defined as their absolute relative speed taken as an average over time. This metric has certain deficiencies: First, it assumes a GPS like scheme for calculation of relative speeds while in a MANET, we cannot assume the existence of GPS, so we have to resort to other techniques for measuring relative mobility. Secondly, it is an "aggregate" mobility metric and does not characterize the local movement of the neighboring nodes to another particular node.

The Reference Point Group Mobility Model (RPGM) proposed in [13] was useful for predictive group mobility management. In RPGM, each group has a logical "center" and the center's motion defines the entire group's motion behavior including location, velocity, acceleration etc.

In [14], They proposed a measure of the network mobility which is relative and depending on neighboring and link state changes. Each node estimates its relative mobility, based on changes of the links in its neighboring. This measure of mobility has no unit, it is independent of any existing mobility models and it is calculated at regular time intervals.

The degree mobility used in [15] was calculated from the change of its neighboring to each node in time. The node mobility degree, represents at a given time $t$ for each node in the ad hoc network, the change variations undergone in its neighboring compared to the previous time $t - 1 - 1$. Thus, nodes that join or/and leave the neighboring of a given node will have surely an influence on the evaluation of its mobility.

. However, the last two measures are not representative's values of a change node's motion with respect to another node.

We can see that none of the metrics described above are suitable for characterizing the relative mobility of nodes in a particular node's neighborhood in a MANET. Hence, we feel that there is a need to develop such a metric which can be used by any routing protocol.

## 4. LOCAL QUANTIFICATION OF NEIGHBOURING MOBILITY

In this section, we define how we estimate nodes mobility in ad hoc network. Mobility is quantified locally and independently of localization of a given node. We represent this local quantification node mobility, by calculating of the distance between a node and its neighbors.

The quantification of the distance can be done using 3 methods:

**Calculate the exact distance**: this is done by two ways:

*1st way: The distance calculation using the GPS*: This operation is done by using a terminal capable of being localized through a positioning system by satellites: GPS. The principle of localization by the GPS system is based on the use of satellite coordinates and the estimation of distances between the receiver satellites. Distances are obtained from the estimation of the TOA (Time Of Arrival) of the signals transmitted by the satellites [16]

*2nd way: Distance calculation function in a simulation environment:* Like NS2, OPNET, tor other simulator.

**Calculate the distance using the RSSI (Received Signal Strength Indication):** in case that the absolute positioning is not accessible, dedicated equipment not available or not possible, in theory, to determine the distance between a transmitter and a receiver we can use the RSSI. RSSI is a generic radio receiver technology metric, which is usually invisible to the user of the device containing the receiver, but is directly known to users of wireless networking of IEEE 802.11 protocol family.

The distance using RSSI can be calculated using the FRIIS transmission formula:

$$P_r = \frac{P_t \ G_t \ G_r \ \lambda^2}{(4\pi)^2 \ d^2 \ L}$$

$P_r$: Receiving power.

$P_t$: Transmitting Power.

$G_t$: Gain of a transmitting antenna = ability to radiate in a particular direction in space.

$G_r$: Gain of a receiving antenna = ability to couple the energy radiated from a direction

in space.

$\lambda$: is the wavelength.

$L$: is system loss factor which has nothing to do with the transmission

$d$: is the distance between the antennas.

Then, to calculate the distance between two nodes that's are equipped by transmitting antennas, the formula is:

$$d = \sqrt{\frac{P_t \ G_t \ G_r \ \lambda^2}{(4\pi)^2 \ L \ P_r}}$$

**Calculate distance using GPS-free** [17]**:** In case that the GPS is not accessible, we can use a GPS-free to localize the neighbor of each node. This method uses a mobile reference to calculate the coordinates of all the nodes in the network. However we can conclude the distance between any nodes. In this part, we use the distance reception power to determinate the distance between the reference and the others nodes.

 To implement this method in AODV protocol, we have to choose the reference:

### ***Choice of the reference***

**a** and **b** are two nodes that want to communicate in a MANET network.

We put **i** the center of the coordination system.

Let **N**: The set of nodes in the network.
   **Pi**: The set of one hop neighbor of the node **i**.
   **dij**: the distance between nodes **i** and **j**.
   **Di**: All distances **dij**
We choose **i /   i ϵ Pa  ,  i ≠  b and I ≠ a**

We choose **p** and **q / p,q  ϵ  Pa , dpq ≠ 0,**

$$\mathbf{p\hat{i}q \neq 180° , p\hat{i}q \neq 0°}$$

Node **i** defines its system of local coordination.

   We set the x-axis as the line **(ip).**

   We set the y-axis as the line **(iq).**

    Thus the system is defined:

The node **i** is the center of the system.

   $$\mathbf{ix = 0     ,    iy = 0}$$

**p** is the node situated on the axis of abscissas

   $$\mathbf{px = dip       ,    py = 0}$$

The node q is located on the ordinate axis

   $$\mathbf{qx = diq \cos \alpha   ,   py = diq \sin\alpha}$$

   with $\mathbf{\alpha = p\hat{i}q}$

 Using the theorem of Al-Kashi:

   $$\mathbf{dpq^2 =  dip^2 + diq^2 - 2 * dip * diq * \cos\alpha}$$

**α** is calculated using this formula:

$$\mathbf{\alpha = \arccos \frac{dip^2 + diq^2 - dpq^2}{2 * dip * diq}}$$

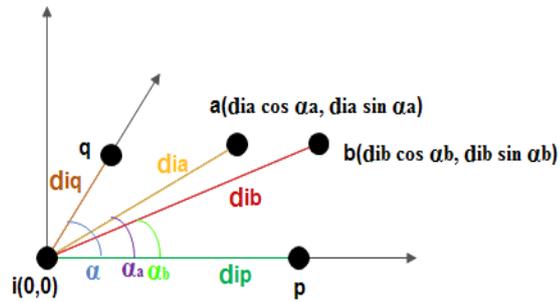

Figure 1 : The reference system

Once the reference is selected, the calculation of the coordinates of the nodes that belong to **Pi** is easy.

$$a \in Pi \begin{cases} a_x = dia \cos \alpha_a \\ \\ a_y = dia \sin \alpha_a \end{cases}$$

$$\alpha_a = a\hat{i}p = \arccos \frac{dip^2 + dia^2 - dpa^2}{2*dip*dia}$$

For **b**, the coordinates are:

$$b \in Pa \begin{cases} b_x = dib \cos \alpha_b \\ dib = dia + dab \\ b_y = dib \sin \alpha_b \end{cases}$$

$$\alpha_b = b\hat{i}p = \arccos \frac{dip^2 + dib^2 - dpb^2}{2*dip*dib}$$

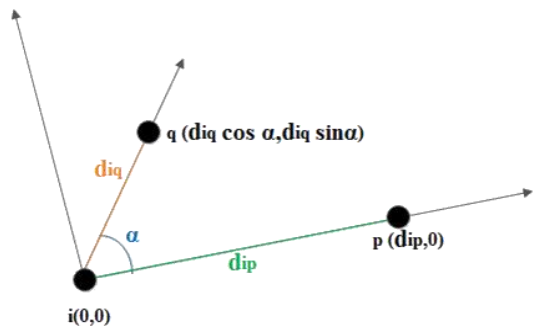

Figure 2: The system of localization local

**Changing the Benchmark:**

We propose in this part to replace the reference by another composed by the **a**'s neighbors having the smallest distances.

However, we choose **i, p, q / I,p,q ∈ Pa , i ≠ b , i ≠ a ,**

$$\mathbf{p\hat{i}q \neq 180°, \ p\hat{i}q \neq 0°}$$

And $\forall \mathbf{k} \in \mathbf{Pi}$ and ∀k ≠ a,b,i,p,q

**dak > dai , dak>dap and dak>daq.**

After the quantification of the distance between all nodes, we can describe the behavior of the node in the network by calculate the average of all the distances **Avg(dij).**

If the average is very high we say that the network nodes are very agitated else the network is supposedly more stable.

## 5. ALGORITHM OF QUANTIFICATION DISTANCE IN THE AODV ROUTING PROTOCOL

In this part, we propose to use one of those methods in the first function of a AODV protocol (rout establishment between a source and a destination).

A node **x** wants to communicate with a node **y**.

**x** diffuse **RREQ**.

Each node receiving **RREQ**, calculates the distance between itself and the neighbor who sent him **RREQ** (in this part we use the exact distance or the distance using the Pr) and broadcasts its table **[neighbors-distance]** to its neighbors.

To use the third method for the quantification of the distance, the algorithm has to change.

A node **x** wants to communicate with a node **y**.

**x** diffuse RREQ.

Each node receiving **RREQ**, calculates the distance between itself and the neighbor who sent him RREQ (in this part we use the exact distance or the distance using the Pr), broadcasts its table **[neighbors-distance]** to its neighbors and choose the reference who has the smallest distance and recalculate the newest distances using the third method.

**N.B.**: the node who receive the **RREQ** is the node a in the previous part.

## 6. SIMULATIONS AND RESULTS

In the following simulations, we applied our proposition to the AODV protocol .For this, we have been used the simulator NS-2 [18], with its implementation of AODV protocol of the version NS-2.35 and Ying-3D [19] to represent some results in 3D.

## 6.1. Environment

The network size considered for our simulations is (1000m×1000m) . The nodes have the same configuration, in particular TCP protocol for the transport layer and Telnet for the application layer. Time for each simulation is of 60s. For each simulation the mobility of the nodes is represented by the choice of an uniform speed between $v_{min} = 0$ and $v_{max} = 100$ m/s. The nodes are moved after a random choice of the new destination without leaving the network (1000m×1000m).

## 6.2. Discussions of results

The results present the local quantification of neighbor's distances during the simulation.

After the application of our proposition on a AODV protocol we obtain:

Table 1: Quantification's results

| RREQ's source | RREQ's destination | Time | Distance | Distance/Pr |
|---|---|---|---|---|
| 7 | 0 | 0,00128055 | 165,397 | 317,654 |
| 8 | 0 | 0,00128093 | 277,895 | 897,138 |
| 3 | 0 | 0,001281 | 299,47 | 1041,43 |
| 17 | 0 | 0,00128117 | 351,078 | 1430,93 |
| 15 | 0 | 0,00128119 | 356,321 | 1474,36 |
| 19 | 0 | 0,00128119 | 358,238 | 1490,13 |
| 11 | 8 | 0,00301544 | 152,84 | 271,179 |
| 9 | 8 | 0,00301572 | 238,948 | 662,809 |
| 1 | 8 | 0,00301579 | 258,118 | 773,428 |
| 7 | 8 | 0,00301585 | 276,525 | 887,669 |
| 0 | 8 | 0,00301585 | 277,731 | 896,079 |
| 15 | 8 | 0,00301586 | 280,446 | 913,022 |
| 10 | 8 | 0,00301598 | 315,33 | 1154,29 |
| 3 | 8 | 0,00301599 | 318,399 | 1176,86 |
| 5 | 8 | 0,00301617 | 371,712 | 1603,97 |
| 19 | 8 | 0,0030162 | 380,623 | 1681,8 |
| 1 | 15 | 0,00584584 | 308,585 | 1105,44 |
| 13 | 15 | 0,00584587 | 316,332 | 1161,63 |
| 5 | 10 | 0,00903948 | 103,368 | 124,039 |
| 4 | 10 | 0,00903976 | 187,032 | 406,083 |
| 9 | 10 | 0,00903978 | 194,023 | 437,009 |
| 11 | 10 | 0,00903982 | 206,903 | 496,956 |
| 1 | 10 | 0,00903988 | 224,54 | 585,286 |
| 18 | 10 | 0,00903999 | 256,821 | 765,673 |
| 16 | 10 | 0,00904017 | 309,608 | 1112,77 |
| 12 | 10 | 0,00904037 | 371,486 | 1602,02 |
| 13 | 10 | 0,0090404 | 378,926 | 1666,83 |
| 19 | 3 | 0,0112602 | 66,4831 | 66,4831 |
| 17 | 3 | 0,0112604 | 136,741 | 217,059 |
| 0 | 3 | 0,011261 | 299,018 | 1038,29 |

In the following figures, we observe the change of the distance between the node 1 and theirs neighbours in the first 3s of the simulation.

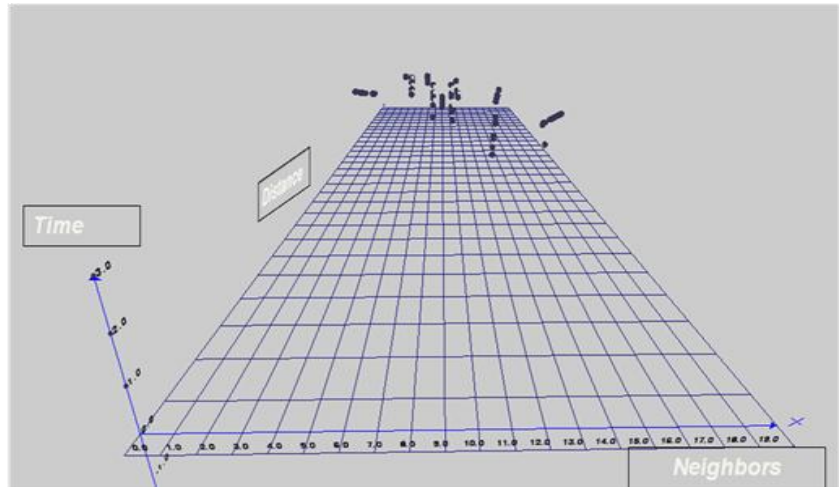

Figure 3: Quantification of the Distance between node 1 and its neighbors during the first 3s of the simulation

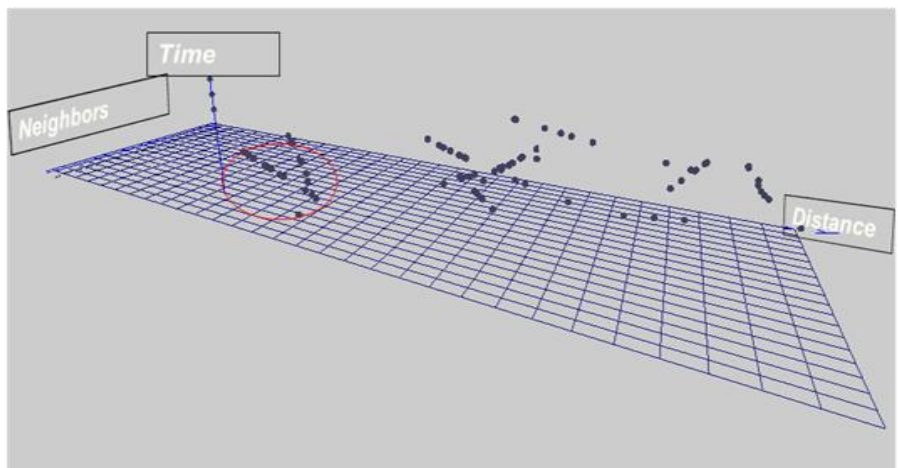

Figure 4: Quantification of the Distance between node 1 and its neighbors during the first 3s of the simulation "Another observation angle"

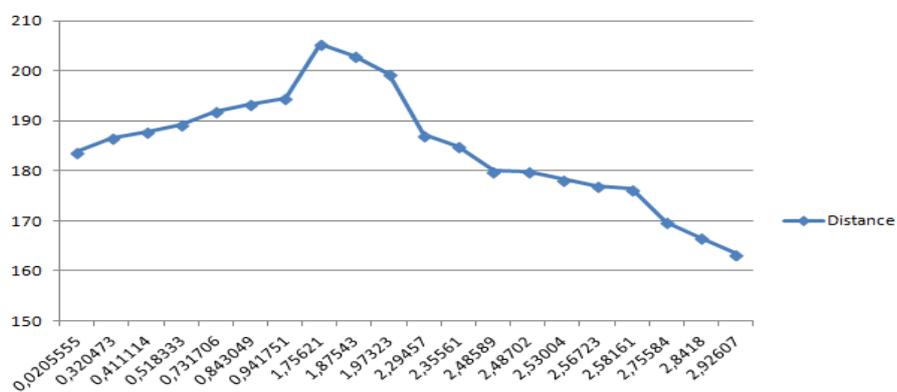

Figure 5: The distance between node 1 and its neighbor node 8 during the first 3s of the simulation

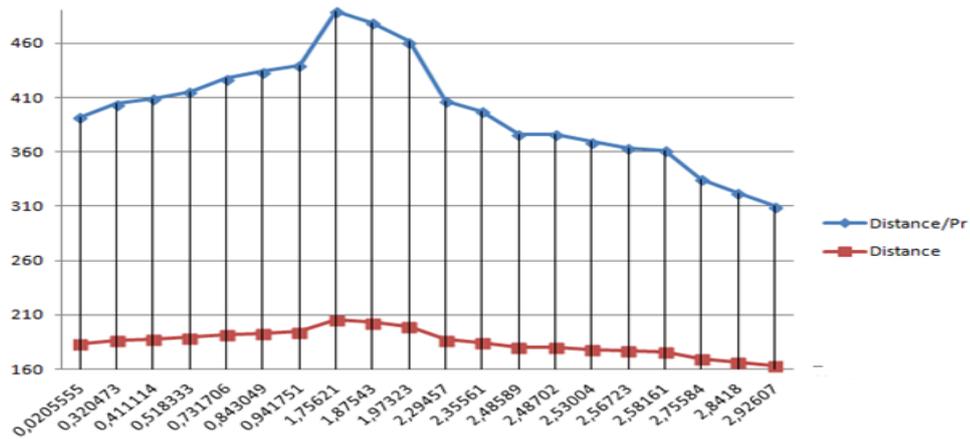

Figure 6: The distance and the distance using RSSI between node 1 and its neighbor node 8 during the first 3s of the simulation

## 7. CONCLUSION:

In this paper, we tried to calculate a local distance between a node and its neighbors in a AODV routing protocol for the ad hoc networks. This metric of mobility that can be used to choose a stable rout to transmit data thus ameliorate the Quality of Service in this kind of networks.

To allow this proposition more really feasible, we present the three methods to calculate the distance between two nodes. First, we use the exact distance with a GPS or using RSSI. In case that the absolute positioning is not accessible, we propose our improved GPS-free implementing in AODV protocol.

**Authors**

**Meryem SAADOUNE**

Received the master degree in Computer Engineering in 2010 from Faculty of Sciences - HASSAN 2 University- Casablanca, Morocco.

Currently, a PhD Student in Computer Science.

Ongoing research interest :

- QoS in Mobile Ad hoc Networks (MANETs) / Wireless Sensor Network (WSN)
- Next Generation Networks

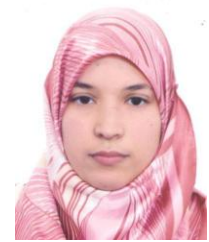

**Abdelmajid HAJAMI**

PhD in informatics and telecommunications,

Mohamed V-Souissi University Rabat-

Morocco. 2011

Ex Trainer in Regional Centre in teaching and

Training

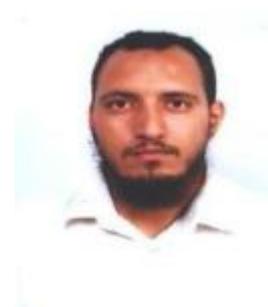

Assistant professor at the Faculty of Science and Technology of Settat in Morocco

Member of LAVETE Lab at Faculty of Science

and Technology of Settat

Research interests:

- Security and QoS in wireless networks
- Radio Access Networks
- Next Generation Networks
- ILE : Informatics Learning Environments

**Hakim ALLALI**

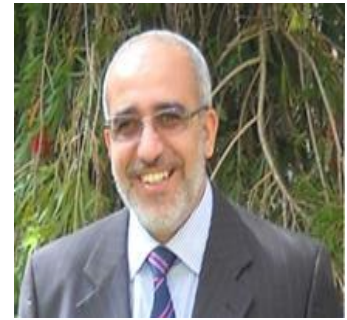

Was born in Morocco on 1966. He received the

Ph.D degree from Claude Bernard Lyon I University (France) in 1993 and the "Docteur

d'Etat" degree from Hassan II-Mohamedia

University, Casablanca (Morocco) in 1997.

He is currently Professor at Faculty of Sciences and Technologies of Hassan 1st University of Settat (Morocco) and director of LAVETE Laboratory.

He is executive manager and founder of IT Learning Campus.

His research interests include technology enhanced learning, modeling, image processing, computer networking and GIS.